# The X/Gamma-ray Imaging Spectrometer (XGIS) on-board THESEUS: Design, main characteristics, and concept of operation

C. Labanti[a], L. Amati[a], F. Frontera[a,b], S. Mereghetti[c], J.L. Gasent-Blesa[m], C. Tenzer[o], P. Orleanski[l], I. Kuvvetli[q], R. Campana[a], F. Fuschino[a], L. Terenzi[a], E. Virgilli[a], G. Morgante[a], M. Orlandini[a], R.C. Butler[a], J.B. Stephen[a], N. Auricchio[a], A. De Rosa[a], V. Da Ronco[a], F. Evangelisti[b], M. Melchiorri[b], S. Squerzanti[b], M. Fiorini[c], G. Bertuccio[d], F. Mele[d], M. Gandola[d], P. Malcovati[e], M. Grassi[e], P. Bellutti[f], G. Borghi[f], F. Ficorella[f], A. Picciotto[f], V. Zanini[f], N. Zorzi[f], E. Demenev[f], I. Rashevskaya[g], A. Rachevski[g], G. Zampa[g], A. Vacchi[h], N. Zampa[h], G. Baldazzi[i], G. La Rosa[j], G. Sottile[j], A. Volpe[k], M. Winkler[l], V. Reglero[m], P. Connell[m], B. Pinazo-Herrero[m], J. Navarro-González[m], P. Rodríguez-Martínez[m], A.J. Castro-Tirado[n], A. Santangelo[o], P. Hedderman[o], P. Lorenzi[p], P. Sarra[p], S. M. Pedersen[q], A. D. Tcherniak[q], C. Guidorzi[b], P. Rosati[b], A. Trois[r], R. Piazzolla[s]

[a]INAF/OAS Bologna, via P. Gobetti 101, I40129 Bologna, Italy;
[b]Università di Ferrara and INFN (Sezione di Ferrara), via Saragat 1, I44122, Ferrara, Italy;
[c]INAF/IASF Milano, via A. Corti 12, I20133 Milano, Italy;
[d]Politecnico di Milano, Dept. of Electronics, Information and Bioengineering (DEIB), Como, Italy and INFN-Milano;
[e]Università di Pavia, Department of Electrical, Computer an Biomedical Engineering, Pavia, Italy;
[f]Fondazione Bruno Kessler, via Sommarive 18, I-38123 Trento, Italy;
[g]INFN (Sezione di Trieste), Padriciano 99, Trieste, Italy;
[h]Università di Udine, Department of Physics;
[i]Università di Bologna, Department of Physics;
[j]INAF – IASF Palermo, Italy;
[k]ASI, Via Del Politecnico snc, Roma, Italy;
[l]Space Research Centre, Polish Academy of Sciences, Bartycka 18A, Warsaw, Poland;
[m]ImageProcessingLaboratory, University of Valencia, c/ Catedrático José Beltrán, 2, E46980, Paterna (Valencia), Spain;
[n]Instituto de Astrofísica de Andalucía (IAA-CSIC), Glorieta de la Astronomía, E18008, Granada, Spain and Dpto. de Ingeniería de Sistemas y Automática, Universidad de Málaga, Avda. Cervantes, 2, E29071 Málaga, Spain
[o]IAAT, Sand 1, D-72076, Tübingen, Germany;
[p]OHB-Italia, Via Gallarate, 150, 20151 Milano, Italy;
[q]DTU Space, Elektrovej Building 327, DK-2800, Kgs. Lyngby, Denmark;
[r]INAF-Osservatorio Astronomico di Cagliari, Via della Scienza 5, 09047 Selargius, Italy;
[s]INAF/IAPS Via del Fosso del Cavaliere, 100, 00133 Roma, Italy.

## ABSTRACT

THESEUS (*Transient High Energy Sky and Early Universe Surveyor*) is one of the three missions selected by ESA as fifth medium class mission (M5) candidates in its Cosmic Vision science program, currently under assessment in a phase A study with a planned launch date in 2032. THESEUS is designed to carry on-board two wide and deep sky monitoring instruments for X/gamma-ray transients detection: a wide-field soft X-ray monitor with imaging capability (Soft X-ray Imager, SXI, 0.3 – 5 keV), a hard X-ray, partially-imaging spectroscopic instrument (X and Gamma Imaging Spectrometer, XGIS, 2 keV – 10 MeV), and an optical/near-IR telescope with both imaging and spectroscopic capability (InfraRed Telescope, IRT, 0.7 – 1.8 μm). The spacecraft will be capable of performing fast repointing of the IRT to the error region provided by the monitors, thus allowing it to detect and localize the transient sources down to a few arcsec accuracy, for immediate identification and redshift determination.

The prime goal of the XGIS will be to detect transient sources, with monitoring timescales down to milliseconds, both independently of, or following, up SXI detections, and identify the sources performing localisation at < 15 arcmin and characterize them over a broad energy band, thus providing also unique clues to their emission physics. The XGIS system consists of two independent but identical coded mask cameras, arranged to cover 2 steradians . The XGIS will exploit an innovative technology coupling Silicon Drift Detectors (SDD) with crystal scintillator bars and a very low-noise distributed front-end electronics (ORION ASICs), which will produce a position sensitive detection plane, with a large effective area over a huge energy band (from soft X-rays to soft gamma-rays) with timing resolution down to a few µs. Here is presented an overview of the XGIS instrument design, its configuration, and capabilities.

# 1. INTRODUCTION

THESEUS (*Transient High Energy Sky and Early Universe Surveyor*) is a mission designed to explore the Early Universe (cosmic dawn and reionization era) and provide a fundamental contribution to multi-messenger and time domain astrophysics [1,2]. By unveiling the bulk of the Gamma-Ray Burst (GRB) population in the first billion years of the Universe, THESEUS will perform studies of the global star formation history of the Universe up to $z \sim 10$ and possibly beyond; detect and study the primordial population III stars, investigate the re-ionization epoch, the interstellar medium (ISM) and the intergalactic medium (IGM) up to $z \sim 8 - 10$ and investigate the properties of the early galaxies determining their star formation properties in the re-ionization era [1, 2, 3]. At the same time, THESEUS will perform a deep monitoring of the X-ray transient Universe in order to locate and identify the electromagnetic counterparts to sources of gravitational radiation and neutrinos; provide real-time triggers and accurate locations of GRBs and high-energy transients for follow-up with next-generation optical-NIR (E-ELT, JWST), radio (SKA), X-rays (ATHENA), TeV (CTA) telescopes.

THESEUS is one of the three M5 missions selected in 2018 by ESA within the Cosmic Vision programme to carry out an assessment phase study and compete for a launch opportunity in 2032. The ESA M5 phase-A study will be completed in spring 2021 with the final down selection to one candidate in summer 2021.

The mission concept includes two X/gamma-ray monitor instruments (SXI and XGIS) with large FOV, an optical/near-IR telescope (IRT) with both imaging and spectroscopic capability, and a data handling unit (DHUs) system capable of detecting, identifying and localizing GRBs and transients in the FOV of the monitors and to request the spacecraft slew in order to bring them in few minutes in the IRT FOV for accurate (few arcsec) location, identification and redshift determination. Furthermore, a Trigger Broadcasting Unit (TBU) onboard THESEUS will provide (within a few tens of seconds) the transmission to ground of the trigger time and position of GRBs and other transients of interest, thus enabling follow-up with the large ground and space observatories.

The *Soft X-ray Imager* (SXI, 0.3 – 5 keV) consists of two X-ray large FOV telescopes based on MicroPore Optics (MPO) in "lobster-eye" configuration, with CMOS detectors in the focus, covering a total FOV of 0.5 sr with 0.5 – 2 arcmin source location accuracy [4].

*The X-Gamma ray Imaging Spectrometer* (XGIS) consists of two coded mask cameras operating in the 2 keV – 10 MeV energy range with spectrometric and timing capabilities and operating also as imager in the 2–150 keV range with source location accuracy <15 arcmin over a FOV of $117 \times 77$ deg$^2$ overlapping the SXI one.

*The Infra-Red Telescope* (IRT) consists of a 0.7 m diameter near infrared telescope, based on optics in a Korsch configuration, Teledyne Hawaii detectors and a filter wheel with filters and dispersive elements. IRT provides imaging and moderate spectroscopic capabilities (R~400) in the 0.7 – 1.8 µm wavelength range over a 15×15 arcmin FOV [5].

The THESEUS mission is planned to be operational for 4 years. This duration is mainly driven by the expected rate of occurrence of high redshift GRBs, which are among the prime targets to be studied. The large fraction of the sky that can be monitored with SXI and XGIS and that is instantaneously accessible by the IRT secure that a sufficient number of these transient sources can be observed within the mission lifetime. At the same time, the nominal duration will allow THESEUS to fully achieve also its main scientific goals for multi-messenger and time-domain astrophysics.

The nominal operational orbit for THESEUS is selected to be a circular low Earth orbit (LEO) with low inclination (<6°). The driver for this choice is fundamentally the low and stable background required for the science performance of the monitors, the XGIS in particular.

Figure 1 shows the THESEUS spacecraft concept in terms of possible spacecraft design and payload accommodation.

This paper describes the XGIS instrument in the configuration that has been optimized for the THESEUS mission throughout the phase-A study.

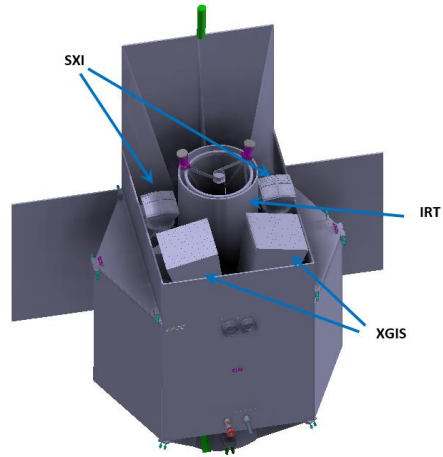

Figure 1 The THESEUS spacecraft concept

## 2. XGIS SCIENTIFIC REQUIREMENTS AND CONCEPT

The inclusion in the payload of a broad field of view hard X/soft gamma-ray detection system, covering a >2sr FOV (that includes the SXI one) and extending the energy band from few keV up to several MeV with at least a few hundreds cm$^2$ effective area over the whole range, will increase significantly the capabilities of the mission. In particular, the monitor will be fundamental for: *a)* detecting and localizing short GRBs, which are a fundamental phenomenon for multi-messenger astrophysics, being up to now the most likely and only one detected electromagnetic counterparts of Gravitational Wave (GW) signals (specifically, from NS-NS and NS-BH mergers), and determining the hard spectrum of these events that make them mostly undetectable with the SXI; *b)* complementing the SXI capabilities for the detection and localization of high-z GRBs, thanks to the large effective area at <10 keV with respect to past and current GRB detectors; *c)* providing unique clues to the physics and geometry of the emission of GRBs and other bright X-ray transients through sensitive timing and spectroscopy over an unprecedented energy band; *d)* detecting possible absorption features in the low-energy spectra of GRBs that may be used for investigating the circum-burst environment, and hence the nature of the progenitor star, as well as inferring the redshift. In addition, as the SXI lobster-eye telescopes can be triggered by several classes of transient phenomena (e.g., flare stars, X-ray bursts, etc), the detection in hard X-rays with the XGIS provides an efficient means of identifying true GRBs. In addition the joint data from the two monitor instruments and the IRT will characterize transient events in terms of luminosity, spectra and timing properties over a broad energy band, thus giving fundamental insights into their physics. See [1, 6] for more details on the scientific case and requirements for the XGIS.

XGIS is an X/γ detector system operates from 2 keV to 10 MeV consisting of:

1. Two identical cameras
2. Two XGIS power Supply Units (XSU)
3. One Data Handling Unit (DHU)

Each camera has a delimited FOV and imaging capabilities in the 2–150 keV range. Above 150 keV the FOV of a XGIS camera is not delimited. With respect to the observation axis of the satellite (X-axis), the two XGIS camera point at two offset directions, ± 20°, so that their FOVs partially overlap and contain the FOV of the other THESEUS instruments (SXI and IRT). In order to meet the top level science goals of the THESEUS mission [6], a set of requirements has been defined for the XGIS camera. These requirements, as well as the performance goals of one camera, are summarized in Table **1**.

| Energy range | 2–150 keV | 150 keV- 10 MeV |
|---|---|---|
| Partially coded FOV (PCFOV) | $77 \times 77$ deg$^2$ | |
| FOV | | $2\pi$ sr |
| Peak effective area | ~500 cm$^2$ | ~1000 cm$^2$ |
| XGIS sensitivity (two combined cameras) | At EOL at least $10^{-8}$ cgs over the 2–30 keV range in 1 s for 99.73% of the observations and at least $3\times10^{-8}$ cgs over the 30–150 keV range in 1 s for 99.73% of the observations | at least $3\times10^{-7}$ cgs over the 150 keV–1 MeV energy range in 1 s for 99.73% of the observations |
| Angular resolution | < 120 arcmin | |
| Source location accuracy | ≤ 15 arcmin 90% confidence level in the 2-150 keV energy band for a source with SNR > 7 | |
| Energy resolution | Better than 1200 eV FWHM @ 6 keV at End of Life | 6 % FWHM @ 500 keV |
| Relative timing accuracy | 7 μs | |

Table 1 XGIS system main characteristics of each XGIS camera

## 3. XGIS INSTRUMENT OVERVIEW

The XGIS system will act both as an imager up to 150 keV, fully overlapping the SXI FOV, and as a spectrometer covering a wide energy range from a few keV to ten MeV, partially overlapping the SXI energy range. The imaging system of the XGIS is based on the coded mask principle, employed successfully in different energy bands on several previous missions such as, e.g. GRANAT/SIGMA, BeppoSAX/WFC, INTEGRAL/JEM-X/IBIS/SPI, RXTE/ASM, Swift/BAT. The mask *shadowgram* is recorded by a position sensitive detector, and can be deconvolved into a sky image. The size of the point spread function in the sky image is determined by the ratio of the mask pixel size and the mask-to-detector distance. The mask pixel size must always be larger than the corresponding detector resolution.

The XGIS detector technology is based on the use of an array of individual Silicon Drift Detectors (SDD) coupled to CsI(Tl) crystal scintillator bars and readout by a distributed low-noise front-end electronics. The SDD are used both for direct detection of low-energy X-rays and as the readout system for the scintillator crystals. A proper electronic technique will distinguish the two kind of signals.

### 3.1 XGIS functional design

Each camera (Figure 2), is composed of a Coded Mask assembly, a Collimator assembly, a Detector assembly with 100 detector Modules organised in 10 Super-Modules and 1 Back-End Electronics board. The overall Camera size is 60×60×91 cm$^3$ for a total weight of about 80 kg, including margin. The camera Modules and Super-Modules are organized to achieve a high level of redundancy. Each camera receives power from an external XGIS Supply Unit (XSU) box and both cameras are interfaced to the XGIS DHU for data transmission and command receival.

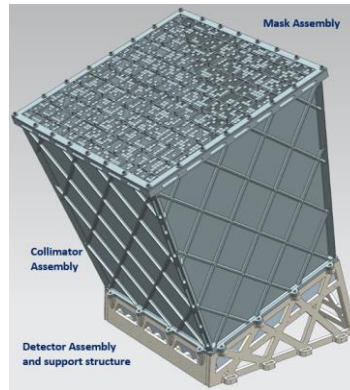

Figure 2 XGIS camera design

Figure 3 shows the functional block diagram for the XGIS. The functions of the main components can be summarized as follows.

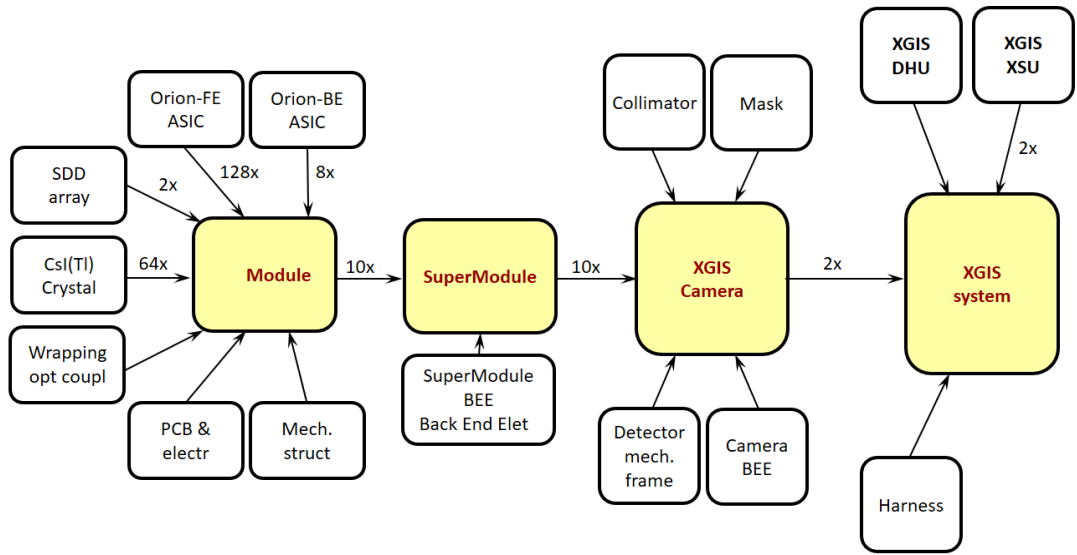

Figure 3 Functional block diagram of the XGIS system, consisting of 2 identical cameras (each one made of 100 Modules organized in 10 Super-Modules), 2 power distribution boxes (XSU) and a Data Handling Unit (in cold redundancy).

The functions of the **Module**, consisting of 64 pixels, are:
- Detect low energy X-ray radiation with SDDs
- Detect γ radiation with the scintillator crystals
- Provide filtered bias voltages to the SDDs and ASICs
- ASICs configuration
- (inside ASICs) Read-out, A/D convert, Time tag the SDD signals
- Provide mechanical support
- Interface mechanically and electrically with the Super-Module

The functions of the **Super-Module**, each consisting of 10 Modules, are:
- Interface mechanically and electrically with the Modules
- Generate bias voltages to the SDDs and ASICs
- Provide logic for modules control and command,
- Provide mass memory for modules data
- Interface the Camera Back End Electronics
- Interface mechanically to the Camera support structure and give stiffness

The main functions of the Camera **Back End Electronics** (BEE) are:
- Provide mass memory for Super-Modules data
- Interface the XSU and DHU
- Interface mechanically to the Camera support structure and give stiffness

The main functions of **XGIS Supply Units** (XSUs) are:
- Supply power to the Super-Modules and Camera BEE
- Interface with DHU

The functions of the **Data Handling Unit** (DHU), described with more detail later, are:
- Operate as Instrument Control Unit

- Interfacing the BEEs and XSUs
- Power distribution
- TC and configuration handling
- On board time management
- Mass memory for storage of telemetry data
- Image integration
- Burst triggering

### 3.2 XGIS accommodation on the spacecraft

The XGIS configuration on THESEUS has been optimized to fulfil the requirement of containing the SXI FOV, and possibly further enlarge the observed sky. For observation strategy reasons, the pointing vector of all THESEUS instruments will be in a band on the sky almost perpendicular to the Sun vector. The FOV of a single unit is determined by the geometry of the camera.

The imaging sensitivity of a single XGIS camera has a complex variation as a function of the position in the field of view, because the "working area" of the detector (i.e. the detector region that measures the flux modulated by the coded mask) depends on the source off-axis angle. In the fully coded field of view (FCFOV), the working area coincides with the whole detection plane, and therefore it has the same dimensions. Therefore, the sensitivity in the FCFOV is approximately constant. For sources outside the FCFOV, the working area is only a fraction of the detector, going linearly to zero with each of the off-axis angles, as shown in Figure 4.

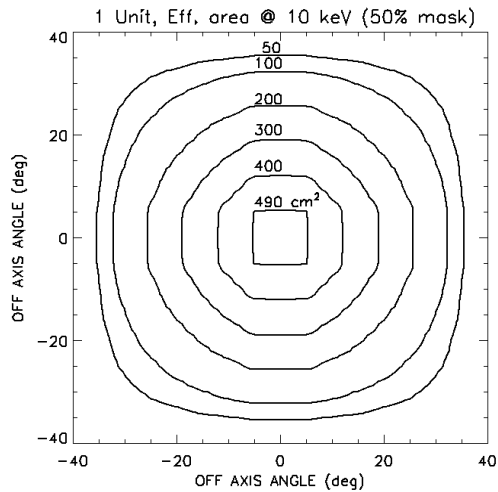

Figure 4 Variation of the effective area of one XGIS camera as a function of off-axis angle

By combining the angular responses of the two units, offset by ±20°, a total 2–150 keV FOV of 117×77 deg$^2$ FWHM is obtained (Figure 5). Details of the XGIS mask and collimator design can be found in [7, 8].

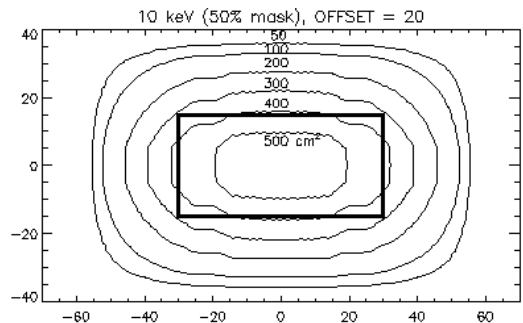

Figure 5 Combined effective area of the two XGIS cameras within the whole FOV. The rectangle at the centre shows the SXI FOV.

Above 150 keV the collimator becomes nearly transparent and the FOV is determined by the passive elements of the satellite. A mass model is being developed also to simulate the response of each camera in this band (Figure 6).

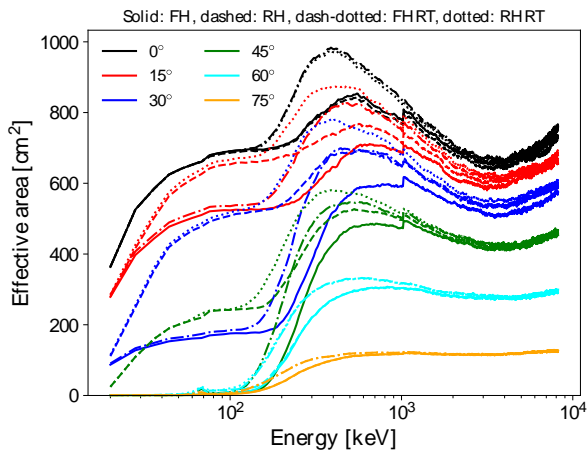

Figure 6 Simulations of the off-axis effective area as a function of the energy, for different solutions in terms of mask thickness and side shielding thickness. Note above ~150 keV that the surrounding structures become transparent and the detector becomes mostly isotropic (response dominated by the cosine effect). FH = Full Height collimator with 0.2 W thickness, RH = Reduced Height collimator (half of FH), RT = Reduced Thickness of the collimator and of the coded mask.

## 4. XGIS ELECTRIC AND DETECTOR ARCHITECTURE

The electrical architecture of the XGIS is shown schematically in Figure 7. It encompasses all electrical subsystems of XGIS and their interfaces with the THESEUS spacecraft platform. From this point of view, the central part of the XGIS is the data handling unit (DHU) with mass memory and power distribution to XSU and containing the instrument control unit (ICU). The burst trigger functionality is implemented as a part of the DHU. The DHU is directly attached to the on-board data handling (OBDH) system. The electrical interface is assumed to be SpaceWire. Each camera back end electronic (BEE) is connected to the DHU for control, data, and monitoring. The bus power is routed through the XSU power distribution unit providing ON/OFF switching and protection capability. XGIS will require about 210 W overall including margin.

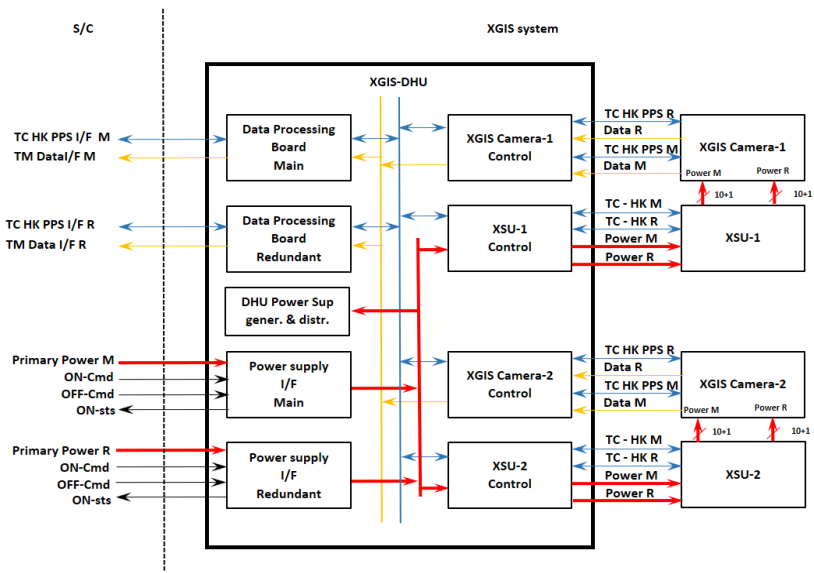

Figure 7 Electrical architecture of XGIS system

## 4.1 The XGIS detection plane

The XGIS detector assembly is represented in Figure 8 with its inclined support structure: the collimator-mask assembly will be fixed on the top of the structure.

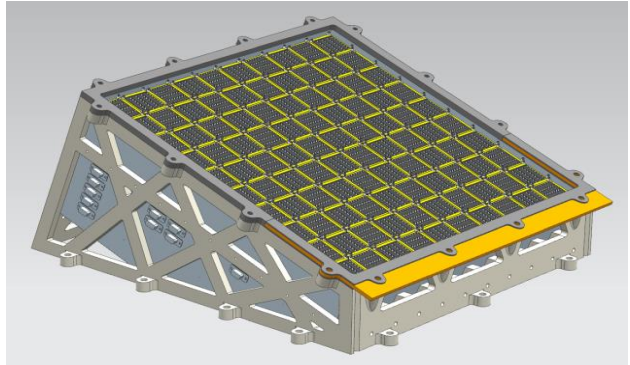

Figure 8 XGIS detector assembly concept

The XGIS detector is organized as follow:
- The basic element is a Module, containing 64 individual pixels, arranged in a monolithic array, for both X and γ-ray detection;
- The XGIS detector plane contains 10×10 modules arranged side by side. A passive space 5 mm wide is interleaved between one Module and the adjacent ones, in this way there are 9 "dead" rows and 9 "dead" columns with width equal to one pixel;
- A XGIS detector assembly has:
    - 6400 SDD pixels for direct detection of X-rays (X-mode, 2–30 keV) with a pitch of 5 mm. Due to the instrument design, part of the SDDs active area is shielded by passive elements, practically resulting in pixels with different 'exposed' areas, 4800 $3.5 \times 5$ mm$^2$ pixels and 1600 $3.5 \times 3.5$ mm$^2$ pixels.
    - 6400 CsI(Tl) pixels for detection of γ-rays (E > 30 keV), have a size $4.5 \times 4.5$ mm$^2$ with a pitch of 5 mm
- The modules are logically and electrically grouped in Super-Modules, each one containing 2×5 Modules and the Super-Module Back End Electronics board (SM-BEE) with common services (Power-Supply and FPGA) in the bottom part of the Super-Module. The Super-Modules will be electrically I/F with a Camera Front End Electronics Board (C-FEE).
- The C-FEE interfaces the XSU for the powering and the Data Handling Unit for Camera commanding, data and HK exchange.

## 4.2 Pixel detector and its operating principle

Figure 9 shows the pixel operation concept. Two Silicon Drift Detectors (SDDs) are placed at the two opposite ends of a scintillator crystal. Low energy radiation (roughly with E < 25–30 keV) is detected in the SDD on the top surface of the Module, while radiation with higher energy reaches the scintillator crystal placed under the SDD. The scintillation light produced by a photon interaction in the crystal is then collected in both SDDs. The crystal is wrapped with a light diffusive foil to enhance light collection.

In XGIS the SDDs have a square cross section 5×5 mm$^2$, whose visible-light entrance window matches the cross section of the scintillator crystal, which is in the form of a small bar 4.5×4.5×30 mm$^3$ in size. The scintillator material is CsI(Tl) with light emission peaking at about 560 nm. While the electron-hole pair creation from X-ray interaction in Silicon generates a fast signal (about 100 ns rise time), the scintillation light collection is dominated by the fluorescent states de-excitation time i.e. 0.68 µs (64%) and 3.34 µs (36%) for CsI(Tl). Therefore a few µs shaping time is needed to avoid a significant ballistic deficit. The discrimination between energy losses in Si and CsI(Tl) can rely on the time coincidence between top and bottom SDDs (occuring in case of scintillation events).

For interactions in Si, the pixel size determines the position resolution in the detector plane. In the CsI, the scintillation light is diffused/reflected on the crystal walls before reaching the SDD; the larger is the scintillation distance from the SDD the greater is its attenuation. By weighting the two SDD signals (top and bottom) the scintillation point along the CsI bar can be evaluated.

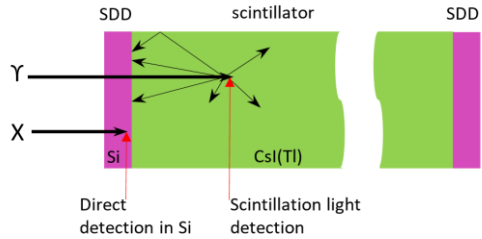

Figure 9 Pixel operation principle

### 4.3 XGIS Module

A detector plane module consists of a monolithic array of 8×8 pixels. The linear size of each pixel is roughly half the size of a single mask element. Figure 10 gives a representation of the module: two set of SDDs (top and bottom) are optically coupled through transparent epoxy and silicone layers to the scintillator crystals. The crystal wrapping optically insulate one crystal to the other ones so that the scintillation light of one crystal can reach only its two coupled SDDs. The SDDs are mounted on a PCB where also the first stages of the electronic chain, the Front End (FE) ASIC, named ORION-FE, one for each SDD, are mounted.

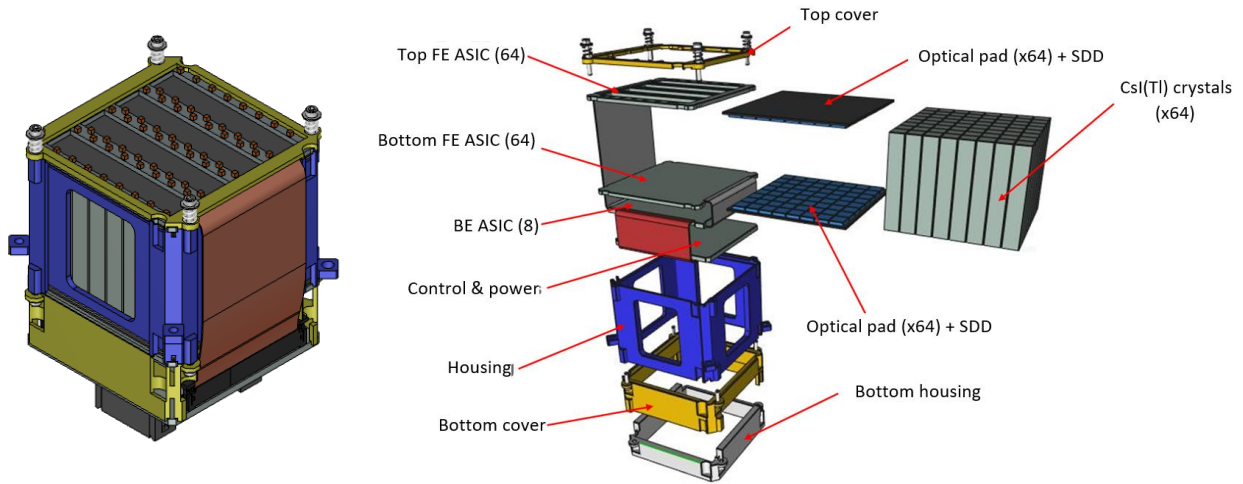

Figure 10 Detector Module design

The front face of the 'top PCB' is exposed to the incoming radiation, while the SDDs are glued on its back face that has large openings for low energy radiation to pass and reach the Si, and it also allows the electrical connection of the SDD anodes to the ORION-FE ASICs glued on the front face of the top PCB and the SDD polarization connections. The electrical signals of the top PCB are sent to the bottom part of the module via a rigid-flex cable.

A 'bottom PCB' has a function specular to the 'top PCB'. In this case the apertures on the PCB have only the function to allow the electrical connection between the SDD anodes and the preamplifiers.

A third board contains the electronics that deals with the signals of the ORION-FE and interfaces with the rest of the system. The main components of this PCB are the Back End (BE) ASICs, named ORION-BE.

The mechanical assembly of the module ensures the correct positioning of the different elements (SDDs, crystals, PCBs) and has been designed to be compatible with vibration loads and with thermo-elastic stress due to the combination of components with different thermal expansion coefficients.

The module design provides a mechanical hard case to which the top and bottom covers are connected. The structures that pack together the components of the module are shown in Figure 11.

In the design the top cover is an open frame that presses the top PCB and thus the top SDD to the optical coupling pads; the pre-load is implemented thanks to the elastic effect of the optical pads and four M1.5 screws.

The bottom cover is directly fixed to the housing without springs and it presses the bottom FEE PCB and thus the bottom SDD to the bottom optical coupling pads.

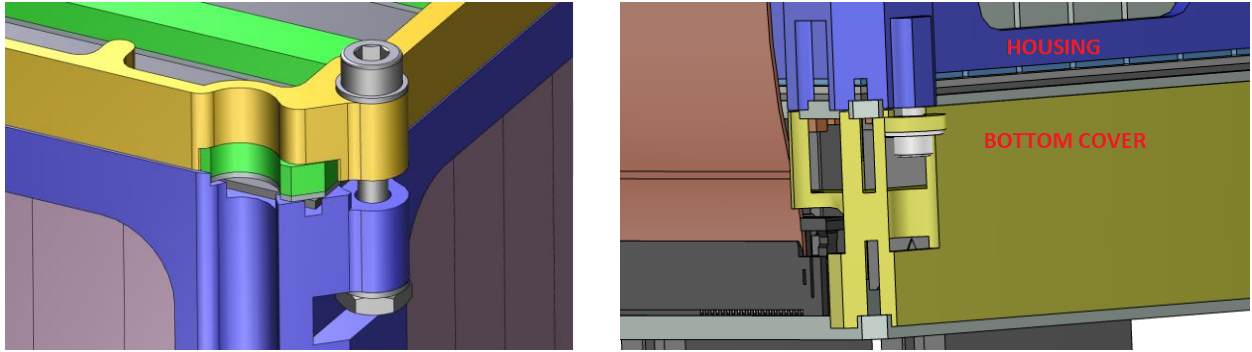

Figure 11 Details of the Top Cover (left) and Bottom Cover (right) connection to the module housing

The housing and the covers are shaped so that the cover screws can be accommodated within 45×45 mm$^2$ foot-print (5 mm pitch, i.e. the same pitch between the SDD centres). The choice of the modular structure is convenient for two reasons: from a technical viewpoint, manufacturing challenges will be confined at Module level as each Module can be tested and qualified independently before being mounted in the Super-Module. It also has positive impact on the program schedule since during detector plane assembly it is possible to easily remove and replace a defective/damaged Module.

### 4.3.1 Silicon Drift Detector

Two arrays of 8×8 SDDs each are employed in each XGIS module (Figure 12), the pitch between the SDD cells is 5 mm. Each array is surrounded by a guard-rings region (~1.2 mm wide). The n-side (anode side) of the SDD will be exposed to the radiation, the p-side will be optically coupled to the scintillators. The sensitive area for X-rays is in principle 5×5 mm, however in practice for each SDD the X-ray detection active area will be smaller being partially obstructed by the top PCB. On the p-side a 0.5 mm wide Al track is deposited between the SDDs for optical separation, therefore the sensitive area for the scintillation light is 4.5×4.5 mm$^2$. The parameters of the SDD array are summarized in Table 2.

| Array size | 42.4 × 42.4 mm$^2$ |
|---|---|
| Si thickness | 450 μm |
| # of SDDs | 64 |
| SDD pitch (n side) | 5 × 5 mm$^2$ |
| Single SDD active area for scintillator (p side) | 4.5 × 4.5 mm$^2$ |
| Metal grid between single SDD (p side) | 0.5 mm wide |
| Typical polarization voltage (one connection) | −150 to −200 V |
| Typical return voltage (one connection/SDD) | −12 to −20 V |
| Single SDD capacitance | 50 fF (typical) |
| Single SDD dark current (typical @ T= 20 °C) | 50 pA (typical) |
| Optical spectral response | 350 – 1000 nm (typical) |
| Quantum efficiency (@ 560 nm) | > 80 % |

Table 2 Main parameters of the SDD array

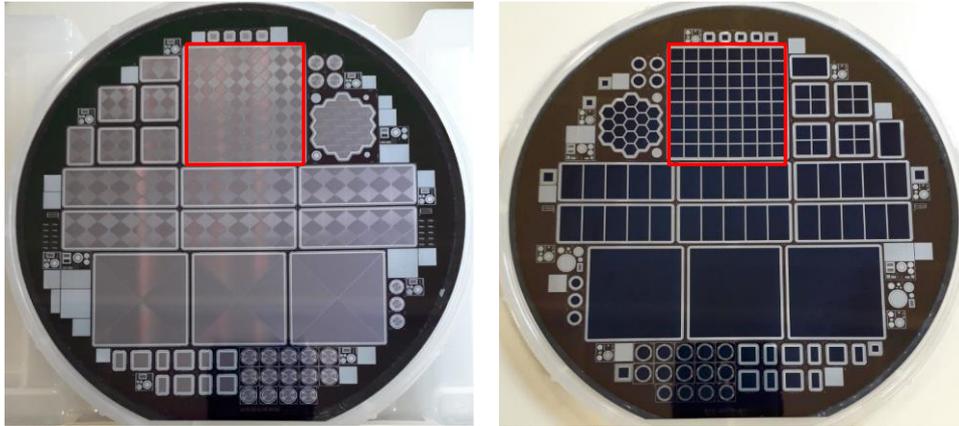

Figure 12 Photos of the two sides of a wafer from the ReDSoX 2019 batch with highlighted the monolithic SDD matrix designed for the XGIS instruments for THESEUS. Left: anode and drift electrodes side (n), input side for X radiation in XGIS. Right: entrance window side (p), input side for the scintillation light.

#### 4.3.2 Scintillator crystal, wrapping and optical coupling

The scintillator crystal is a CsI(Tl) bar of 4.5×4.5×30 mm$^3$ in size. The scintillator bars will be polished on the sides in contact with the SDDs and rough on the other surfaces. Each crystal will be wrapped on all sides except for the SDD-coupled ones with a diffusive/reflective coating, 0.15–0.2 mm thickness. This will guide the scintillation light into the photodetector. Each CsI(Tl) crystal will be optically coupled to the SDDs by a silicone pad transparent and flexible, less than 1 mm thick.

#### 4.3.3 The ORION-ASICs

The signals from an SDD will be collected and processed by the ORION ASICs chipset (ORION-FE and ORION-BE) [8,9], with a structure as shown Figure 13.

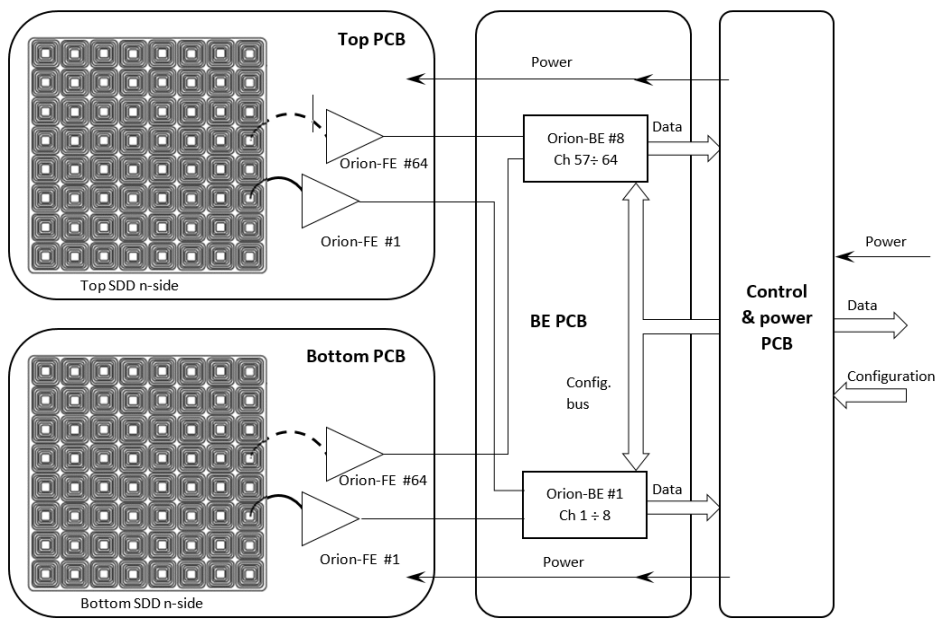

**Figure 13 Module electronic design based on the ORION-FE and ORION-BE ASICs**

A single channel ORION-FE ASIC is placed near the SDD anode, in order to maintain the stray capacitance as low as possible. The ORION-FE ASIC collects the SDD charge signal and performs a pre-amplification-filtering and transmits the processed signal to the ORION-BE ASIC placed few cm away on the bottom of the module in the BE-PCB. In order to avoid inter-channel crosstalk, a current-mode signal transmission is implemented. The functions operated in the ORION-BE are represented in Figure 14; : a three-shapers processing for X and γ originated signals, followed by three 12 bits A/D converters. As a baseline, one ORION-BE will operate on 8 pixels (16 SDDs), processing signals from 8 ORION-FE.

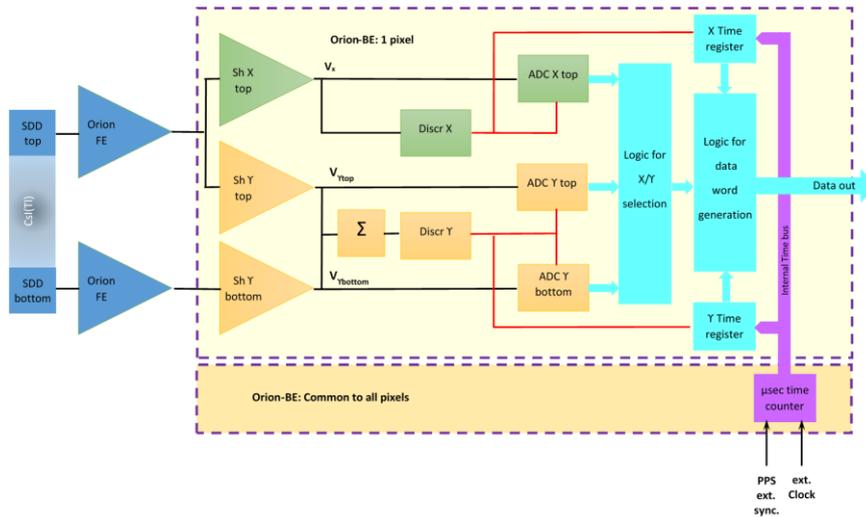

Figure 14 Architecture of a channel of the ORION-BE ASIC: as a baseline, one ORION-BE will operate 8 pixels

The rationale for this architecture is the following:
- An X-ray (2–30 keV) will be detected only in the top SDD, in this case, the few photons in this energy range reaching the scintillator generate signals below the threshold levels and will not be detected:
    o The rise time of the signal will be fast (a few hundreds of ns).
    o As the X-ray detection is (almost) point-like in the SDD, the arrival time of the signal at SDD anodes will be eventually delayed up to 1 μs with respect to the event occurrence due to the charge drift time in the SDD; the time marking of the event is then affected both by the uncertainly due to jitter of the trigger and by the unknown position of the photon interaction in the Si.
    o The best signal/noise ratio will be achieved with a short shaping time (1 μs)
    o The signal discrimination will operate only on the top SDD signal.
- A γ-ray (>20 keV) will be detected in time coincidence in both top and bottom SDDs, in this case:
    o The signals will be slow (of the order of few μs, due to the scintillation light characteristic timescale).
    o The scintillation light will involve the whole SDD surface, so that the maximum drift time of the charge should be taken into account to avoid ballistic deficit.
    o The best signal/noise ratio will be achieved with a shaping time of the order of 3 μs (typical)
    o The best discrimination will be achieved by operating on the sum of top and bottom SDD signals.

Also, mixed X and γ-ray events in time coincidence can occur. A triggered event data will be assembled in ORION-BE as a data word with (number of bit in bracket):

| Time (24) | Address (3) | Int. Service (12) | X/γ (1) | SDD Bottom (12) | SDD Top (12) |

Table 3 Event data format

The Time resolution of an event will be of the order of 1 μs; it will be defined using the Pulse Per Second (PPS) and Ext Clock signals. The PPS will have a frequency of 1 Hz and will be synchronized with the satellite time.

The SDD Bottom value is used to determine the type of event. When SDD bottom is 0 (or < TBD value) the event will be considered an X otherwise it will be considered γ or mixed and the values inserted by ORION-BE in the data word are:

| Event type | Time | X/γ | SDD Bottom | SDD Top |
|---|---|---|---|---|
| X | X-Time | 1 | 0 | $ADC_{X,top}$ |
| γ | γ-Time | 0 | $ADC_{Y,bottom}$ | $ADC_{Y,top}$ |

Table 4 Data type produced by the ORION-BE ASIC

Prototypes of ORION family ASICs are under development during THESEUS phase-A, as they are an evolution of the LYRA family ASICs already produced and used in the HERMES nanosat program [10, 11,12]. The main parameters of the ASIC and of the embedded ADC are reported in Table 5.

| Parameter | Value | |
|---|---|---|
| | X-Photons | γ-Photons |
| Number of Channels | ORIONFE=1 SDD, ORIONBE=8pixels (i.e. 16 SDD) | |
| Temperature Range | -30 …. + 30 °C | |
| Input Range | 20000 e- (37 keV in Si) | 360 000 e- 10 MeV (in CsI(Tl)) |
| Event Rate | 10 …. 100 Events/sec | |
| Shaping Time | 0.5 .... 1 μs | 3 μs |
| Detector Leakage Current range | ≤ 1nA | |
| Detector Capacitance | ≈ 50 fF | |
| Noise @ T = -20 °C | 13 e- rms (112 eV FWHM in Si) | 33 e- rms (4.3 keV FWHM in CsI(Tl)) |
| Linearity | ± 1% | |
| Power Consumption | 1.9 mW/channel | |

ADC architecture → Second-order, single-bit incremental ADC
Resolution → 12 bits
Clock frequency → $f_s$ = 10 MHz
Number of clock cycles per conversion → OSR = 256 (programmable 128, 256, 512)
Conversion time → $T_c = OSR/f_s$ = 25.6 μs
Maximum event rate allowed → 39000 events/s

Table 5 Main characteristics of the ORION ASICs (top) and details of the embedded ADC

### 4.4 Super-Module and Camera - Back End Electronics

An XGIS Super-Module is a logical and hardware subset of the detector assembly and consists of 10 modules, a mechanical grid for assembling the Modules and a Super-Module Back End Electronics board (SM-BEE) mounted in the lower part of the frame containing common services (Power-Supply and FPGA). The SM-BEE functions are:
- Provides the Power Supply to 10 Modules
- Commands the Modules
- Collect the data (events/HKs) from the Modules
- Handle HKs and alarms from the Modules
- Interface all the functions (commands, HKs, Alarms, Data) with the XGIS Camera Back End Electronics

The architecture of the SM-BEE will be built around one FPGA that will control 80 ORION-BE ASICs.

The functions of the Camera Back End Electronics (CBEE) are:

- Power supply I/F with the XSU
- Power supply distribution and management of the Super-Modules
- Data I/F with the Super-Modules
- Data buffering and I/F with XGIS-DHU
- TLC I/F with XGIS-DHU
- TLC implementation and verification
- HK management and transmission to XGIS-DHU
- Alert management

The CBEE board will host two 'sections' in cold redundancy each connected to the DHU main and redundant section and with the 10 Super-Modules.

### 4.5 The XGIS Supply Unit (XSU)

The modular structure of XGIS Camera will be maintained in the design of the power supply that will be split into two XSU boxes each one powering one camera and hosting two 'sections' in cold redundancy each connected to the DHU main and redundant section. Each XSU will
- supply the different Super-Modules with dedicated floating grounds, with Main and Redundant lines in cold redundancy
- provide EMC filtering, primary on/off (if needed), interface to DHUs with HK monitoring;
- XGIS Supply Unit consumes an input primary power of 78.7 W, delivers an output secondary power of 49.4 W, for an efficiency of 62,8%. The relatively small efficiency is due to worst case specifications of the commercially available, space graded DC-DC modules dedicated for low voltage outputs;
- overall dimensions of XSU Unit are approximately 116 mm × 320 mm × 200 mm (H×W×D), dimensions of the legs are not included. The very flat shape is due to the relative big power dissipation and necessity to transfer the heat to S/C through the bottom area of the unit.

The voltage for SDD polarization (between –150V to –200V) will be generated at Super-Module level, in the SM-BEE board

## 5. XGIS MECHANICAL AND THERMAL DESIGN

The mechanical design of the XGIS camera is driven by the need to establish and maintain the alignment of the coded mask and the detector plane. This is essential to achieve stable and accurate imaging properties of the system. The thermal design of the XGIS is also driven by this concern, as well as the need to maintain the SDDs operative temperature in the 0- 40 °C range. An overview of the main camera components is shown in Figure 2.

### 5.1 Coded mask and collimator

The 1 mm thick Tungsten coded mask (described in [8]) has an envelope of 600×600 mm$^2$ with a self-supporting pattern while guaranteeing the maximum transparency. The coded mask is mounted between an upper and a lower grid connected to an external frame that allows its mechanical assembly on the collimator.
The Collimator Assembly, made of Al alloy 1 mm thick, has two main objectives.I It provides the necessary stiffness to the structure connecting the Coded Mask Assembly with the detector assembly, and it accommodates 0.25 mm thick tungsten slabs that act as a lateral passive shield for the XGIS Imaging System.
The combination of the coded area with the collimator aperture provides a FOV of 77.0×77.0 deg$^2$ up to 150 keV.

### 5.2 Detector structure

A Super-Module frame, see Figure 15 will position and hold 10 modules anchoring each module housing as detailed in Figure 16, The detector plane will be assembled joining together 10 Super-Modules.

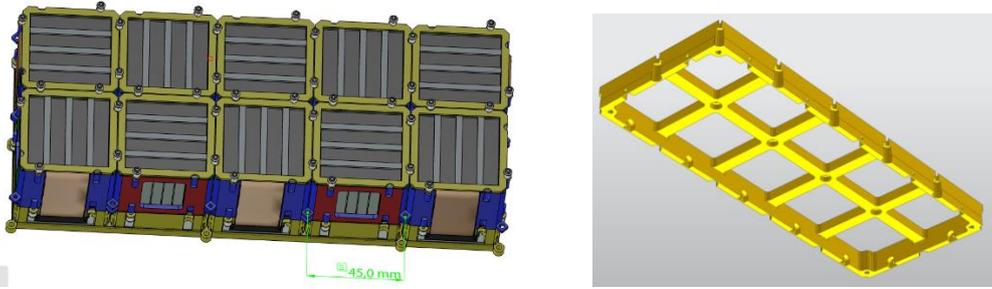

Figure 15 Left: Super-Module arrangement (view from the radiation entrance side); right Super-Module frame (bottom view, the Modules are mounted on the upper part of the frame, and screwed together via the vertical holes than can be seen in the central rib and in the periphery of the frame, the horizontal holes will be used to connect the frame together)

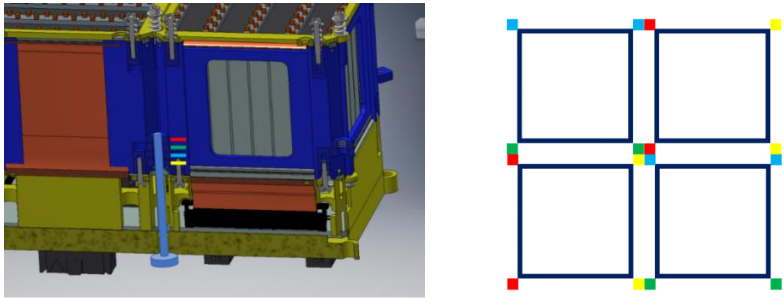

Figure 16 Details of the mechanical assembly of the DM (left), and scheme of DM alignment through using the slots placed at different levels in the housing of the module (right)

The Super-Module frames will be assembled together to form an unique structure (see Figure 17) screwing one to each other using the holes on the perimeter of the frame; a reinforcement element will ensure the stiffness of the structure. The Camera Back End Electronics PCB, serving all the Super-Modules, will be mounted below this frame on the opposite side of the Super-Modules. The whole Super-Module structure will be fixed to the external Camera Support Structure (Figure 8) that allows each XGIS camera to point with 20° offset from the X-axis of the satellite and that also supports the collimator and mask. A box cover will obscure and shield the detector plane.

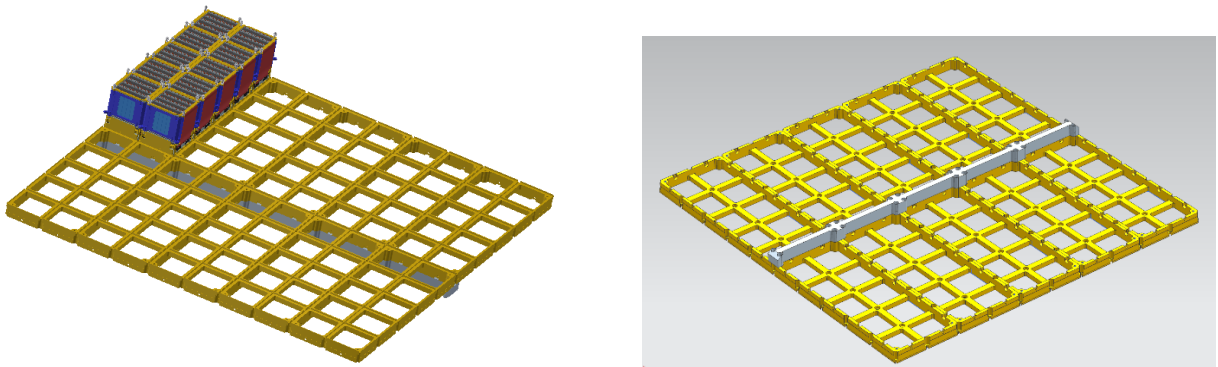

Figure 17 Top (left) and Bottom (right) view of the Super-Module frames screwed together to form the structure of the whole detector module assembly and with a reinforced element to increase its stiffness

### 5.3 XGIS thermal design

A schematic of the heat transfer processes for each XGIS camera is shown in Figure 18.

The elements considered in the thermal control design are the coded mask, the collimator and the detector. The latter is thermally described as a structure divided into layers representing the PCBs, notably the top and bottom PCB where the SDD are placed, and the interleaving material like the CsI scintillators. Thermal conductive connections are used to simulate the heat exchange between the PCBs occurring through the flex harness and from the metal grid of the Super-Module scheme.

The main thermal interfaces with the S/C are a Cold Finger flange connecting the top PCB-SDD layer to the heat pipes coming from the spacecraft radiator dedicated to XGIS and the base of the Titanium support frame bolted to the payload platform.

A set of analysis cases (cold and hot steady, cold transient) have been run in order to verify the consistence of the outputs in terms of temperature and heat exchange of the main interfaces and reference points. The models have been integrated in the spacecraft model and two orbital cases have been defined and run as cold and hot radiative cases with the following common parameters:

- Circular orbit. Altitude: 600 km
- Inclination: 5.4°
- Pitch: –30°

The position of the spacecraft for the steady state cold case is that reached in the middle of the eclipse period when the S/C is on the opposite side of the Earth w.r.t. the Sun and the payload is exposed to the cold space. The position of the spacecraft for the steady state hot case is directly lit by the Sun and the payload is exposed to the Earth surface. The results are summarised in Table 6

The temperatures are compliant with the requirements; this is especially relevant for the top SDD layer at the End of Life of the mission due to the increase in the SDD bulk leakage current due to radiation damage [13].

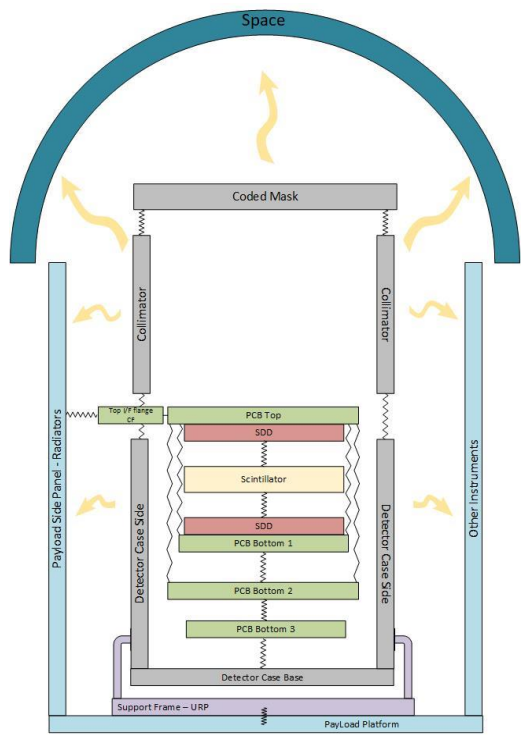

Figure 18 The thermal scheme of the XGIS single camera

| Instrument part | Cold Case T [°C] | Hot Case T [°C] |
|---|---|---|
| Top PCB | 8.2 | 9.6 |
| Top SDD | 8.3 | 9.8 |
| Bottom SDD | 10.1 | 11.7 |
| PCB bottom 1 | 10.3 | 11.8 |
| PCB bottom 2 | 15.9 | 18.0 |
| PCB bottom 3a | 16.7 | 18.9 |
| PCB bottom 3b | 21.2 | 24.7 |
| Detector case side | 12.7 | 17.6 |
| Detector case base | 21.1 | 24.7 |
| Collimator | -20.5 | -4.1 |
| Coded mask | -29.6 | -7.7 |

**Table 6 Temperature distribution of the XGIS Camera in the cold and hot cases with the Cold Finger temperature set 5 °C and the S/C platform at 20 °C.**

# 6. XGIS DATA AND TELEMETRY

In its Observation mode XGIS fulfils the Theseus Science driven modes namely:
- Survey ("burst hunting") mode: monitoring the sky with the wide-field monitors waiting for GRBs to occur.
- Burst mode when a GRB is detected the satellite will autonomously fast repoint to place the position of the GRB in the field of view of the IRT.
- IRT Follow-up mode: IRT observational sequence of a GRB brought in its FOV.
- External trigger mode: when THESEUS will observe a transient discovered by other facilities
- IRT observatory mode: when IRT is used as an observatory for pre-selected targets through a GO program

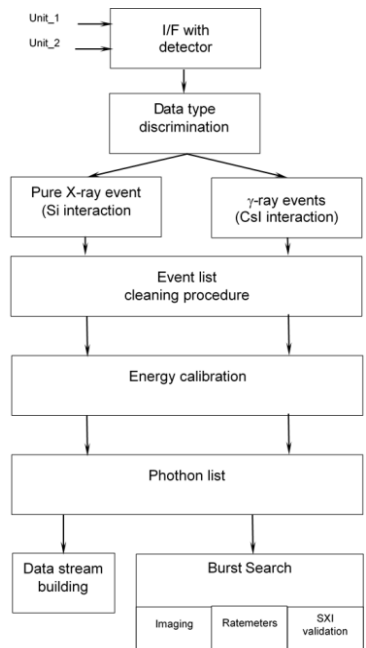

Figure 19 XGIS data task flow in the DHU

XGIS data will be used to produce or validate triggers and will be organized in images and photon by photon streams with different characteristics depending on the THESEUS mode. During passage through the South Atlantic Anomaly XGIS data will be discarded.

The XGIS overall data flow is illustrated by the diagram in Figure 19. The digital signals from the ASICs are stored at different levels in the Super-Module and in the Camera Back End Electronics and then sent to the DHU. Each event is recorded with the information of Table 3; as event time tagging is done at ASIC level the stream of data arriving to DHU is not ordered in time.

### 6.1 Pre-processing of raw data

Before being used for trigger generation and 'data stream' building, XGIS raw data will be:
- Sorted and ordered in time
- Processed in order to distinguish between X and γ events, on the basis of the selection made at ASIC level.
- For X events, calibrated in energy using proper SDD look up tables.
- For γ events, calibrated in energy using look up tables, and used to evaluate the depth of interaction in the detector.
- Cleaned for particles events on the basis of the number of time coincident triggered pixels, their topology and their energy deposits.
- Cleaned to improve signal/background ratio rejecting low energy events occurring on the S/C side.

### 6.2 GRB trigger search

Essentially, two methods will be used to search for GRBs in XGIS data based on rate and/or on images variations.

The rate trigger method, which operates on the whole XGIS energy range, will compare variation of count rate on different energy bands, different integration times and different detector areas. Each estimated rate will be compared with the background rate evaluated dynamically around the orbit.

The image analysis operating up to 150 keV will compare images integrated on different time intervals (e.g. 20 s, 100 s, 500 s and only one or two energy ranges) with background images evaluated dynamically around the orbit with the same parameters.

### 6.3 Qualification of the SXI triggers

When a transient signal triggers a detection in SXI, while no trigger occurs in XGIS (due to an event signal below the one hat produces an automatic trigger), a qualification procedure is performed looking for the corresponding event in XGIS (both rate and imaging).

### 6.4 Data stream building and telemetry

The XGIS data will be organised in different telemetry packets with the following main characteristics:
- Photon by Photon Full Energy data Stream in the full energy range (> 2 keV) cleaned from particles.
- XGIS Photon by Photon High Energy data Stream in the energy range > 30 keV (γ events) cleaned from particles.
- XGIS Histogram and Energy data Stream in the energy range between 2–150 keV i.e. histograms pixel by pixel and in 8 energy channels integrated in e.g. 32 s of the two XGIS Camera and light curves ratemeters on the whole XGIS camera in 4 X energy bands integrated for 16 ms.

The expected telemetry rate, in Gbit/orbit, in the main THESEUS modes is summarized in Table 7.

|  | Imaging | Photon by photon | Total |
|---|---|---|---|
| Survey mode | 0.2[+] | 2.5[*] | 2.7 |
| Follow up mode | - | 0.3[**] | 0.3 |
| Burst mode | - | 3.6[**] | 3.6 |

Table 7 Estimated XGIS telemetry load in the main THESUS modes ([+] 2 < E < 30 keV, [*] E > 30 keV, [**] Full energy range)

## CONCLUSIONS

The XGIS instrument onboard THESEUS mission has been presented. This work includes a summary of the main requirements of the instrument as well as of its main characteristics. The XGIS instrument parts have been also described, two Cameras (Detector Unit), two Power Supply Units and a Data Handling Unit. Each Camera is divided into an Imaging System and a Detector Plane.

The innovative and challenging design of the XGIS detector plane, which use Silicon Drift Detectors and Crystal Scintillator bars technologies, are included herein. This design covers an operating energy range of 2 keV – 10 MeV.

Moreover the instrument includes imaging capabilities for X-gamma ray radiation, up to 150 keV based on signal multiplexing technique using a coded mask at each Camera.

In addition to the electric and detector architecture and the mechanical and thermal design of the instrument, an approach to the Data and Telemetry assessment based on the concept of the XGIS operation at this stage is also presented.

The design of the instrument complies with the requirements at this stage of the project.

## ACKNOWLEDGEMENTS

The Phase A study of the THESEUS/XGIS instrument is supported by ASI-INAF Agreement n. 2018-29-HH.0, OHB Italia/ - INAF-OASBo Agreement n.2331/2020/01, by the European Space Agency ESA through the M5/NPMC Programme and by the AHEAD2020 project funded by UE through H2020-INFRAIA-2018-2020.
By the Spanish Ministerio de Ciencia e Innovación, PID2019-109269RB-C41.
By Polish National Science Center, Project 2019/35/B/ST9/03944 and Foundation for Polish Science, Project POIR.04.04.00-00-5C65/17-00,

## REFERENCES

[1] Amati, L., et al., "The THESEUS space mission concept: science case, design and expected performances", Advances in Space Research, Vol.62 (1), 191-244 (2018).

[2] Stratta, G., et al., "THESEUS: A key space mission concept for Multi-Messenger Astrophysics", Advances in Space Research, Vol.62 (3), 662-682.

[3] Amati, L., et al., "The Transient High-Energy Sky and Early Universe Surveyor (THESEUS)" Paper 11444-302, this Conference

[4] O'Brien, P., et al., "The soft x-ray imager on THESEUS: The transient high energy survey and early universe surveyor", Paper 11444-304 this Conference

[5] Götz, D., et al., "The Infra-red Telescope on board the THESEUS mission", Paper 11444-305, this Conference

[6] Amati, L., et al., "The X/Gamma-rays Imaging Spectrometer (XGIS) on-board THESEUS: Science case, requirements, concept, and expected performances", Paper 11444-279, this Conference

[7] Mereghetti, S., et al, "Scientific simulations and optimization of the XGIS instrument on board THESEUS", Paper 11444-276, this Conference

[8] Gasent-Blesa, J.L., et al., "The XGIS imaging system on board the THESEUS mission", Paper 11444-278, this Conference

[9] Mele, F., et al., "The ORION Chipset for the X-Gamma Imaging Spectrometer Onboard of the THESEUS Space Mission", IEEE-2020 Nuclear Science Symposium and Medical Imaging Conference Proceedings (NSS/MIC), Boston, US-MA, 2020 (to be published).

[10] Fuschino, F., et al., "The XGIS instrument on-board THESEUS: The detection plane and on-board electronics", Paper 11444-277, this conference.

[11] Gandola, M., et al., "LYRA: A Multi-Chip ASIC Designed for HERMES X and Gamma Ray Detector", IEEE-2019 Nuclear Science Symposium and Medical Imaging Conference Proceedings (NSS/MIC) Manchester UK.

[12] Fuschino, F., et al., "An innovative architecture for wide band transient monitor on board HERMES nano-satellite constellations", Paper 11444-167, this conference.

[13] Campana, R., et al.," Background simulations for the Large Area Detector onboard LOFT " (2013), Exp. Astr. 36,451